\input harvmac
\input tables
\def\Title#1#2{\rightline{#1}\ifx\answ\bigans\nopagenumbers\pageno0\vskip1in
\else\pageno1\vskip.8in\fi \centerline{\titlefont #2}\vskip .5in}

%
\font\ticp=cmcsc10

\def\ajou#1&#2(#3){\ \sl#1\bf#2\rm(#3)}
\def\jou#1&#2(#3){\unskip, \sl#1\bf#2\rm(19#3)}

\def\frac#1#2{{#1 \over #2}}

\def\thta#1#2{\Theta_#1(i\epsilon_#2 t|i t)}

\lref\hjs{H. Sheinblatt, 
``Statistical Entropy of an Extremal Black Hole with
0- and 6-Brane Charge'',
hep-th/9705054}
\lref\mrd{M. Douglas, ``Superstring dualities, Dirichlet
branes and the small structure of space'', 
hep-th/9610041.}
\lref\dps{M. Douglas, J. Polchinski, and A. Strominger, 
``Probing Five-Dimensional Black Holes with D-branes'',
hep-th/9703031.}
\lref\bfss{T. Banks, W. Fischler, S. Shenker, and L.Susskind, 
``M Theory as a Matrix Model: A Conjecture'',
\ajou  Phys. Rev. &D55 (1997) 5112, 
hep-th/9610043.} 
\lref\mbranes{T. Banks, N. Seiberg and S. Shenker, 
``Branes from Matrices'',
\ajou Nucl. Phys. &B490 (1997) 91,
hep-th/9612157.}
\lref\tasi{J. Polchinski, 
``TASI Lectures on D-Branes'',
hep-th/9611050.}
\lref\acny{A. Abouelsaood, C. Callan, C. Nappi, and S. Yost,
\ajou Nucl. Phys. &B280 (1987) 599.}
\lref\pair{C. Bachas and M. Porrati, 
``Pair Creation of Open Strings in an Electric Field'',
\ajou Phys. Lett. &B296 (1992) 77,
hep-th/9209032.}
\lref\bachas{C. Bachas, 
``D-Brane Dynamics'',
\ajou Phys. Lett. &B374 (1996) 37,
hep-th/9511043.}
\lref\lifsone{G. Lifschytz, 
``Comparing D-branes to Black-branes'',
hep-th/9604156.}
\lref\jabbari{H. Arfaei and M. M. Sheikh Jabbari,
``Different D-brane Interactions'',
\ajou Phys. Lett. &B394 (1997) 288,
hep-th/9608167.}
\lref\lifstwo{G. Lifschytz, 
``Probing Bound States of D-branes'',
\ajou Phys. Lett. &B388 (1996) 720,
hep-th/9610125.}
\lref\oferberk{O. Aharony and M. Berkooz, 
``Membrane Dynamics in M(atrix) Theory'',
\ajou Nucl.Phys. &B491 (1997) 184,
hep-th/9611215.}
\lref\lifsmath{G. Lifschytz and S. Mathur, 
``Supersymmetry and Membrane Interactions in M(atrix) Theory'',
hep-th/9612087.}
\lref\lifsfour{G. Lifschytz, 
``Four-brane and Six-brane Interactions in M(atrix) Theory'',
hep-th/9612223.}
\lref\vjfinn{V. Balasubramanian and F. Larsen, 
``Relativistic Brane Scattering'',
hep-th/9703039.}
\lref\cheptsey{I. Chepelev and A. Tseytlin, 
``Long-distance interactions of D-brane bound states 
and longitudinal 5-brane in M(atrix) theory'',
hep-th/9704127.}
\lref\noforce{A. Tseytlin,
```No force' condition and BPS combinations of p-branes
in 11 and 10 dimensions'',
hep-th/9609212.}
\lref\juan{J. Maldacena, 
``Probing near extremal black holes with D-branes'',
hep-th/9705053.}
\lref\juanphd{J. Maldacena, 
``Black Holes in String Theory'',
Ph. D. Thesis, Princeton University, June 1996,
hep-th/9607235;
J.Maldacena,
``Black Holes and D-branes'',
hep-th/9705078.}
\lref\wati{W. Taylor, 
``Adhering 0-branes to 6-branes and 8-branes'',
hep-th/9705116.}
\lref\dkps{M. Douglas, D. Kabat, P. Pouliot, and S. Shenker,
``D-branes and Short Distances in String Theory'',
\ajou Nucl.Phys. &B485 (1997) 85,
hep-th/9608024.}
\lref\SandV{A. Strominger and C. Vafa,
``Microscopic Origin of the Bekenstein-Hawking Entropy'',
\ajou Phys. Lett. &B379 (1996) 99,
hep-th/9601029.}
\lref\VandK{E. Keski-Vakkuri and P. Kraus,
``Notes on Branes in Matrix Theory'',
hep-th/9706196.}
\lref\antibrane{T. Banks and  L. Susskind,
``Brane - Anti-Brane Forces'',
hep-th/9511194.}
\lref\hp{G. Horowitz and J. Polchinski, hep-th/9612146.}
\noblackbox
\Title{\vbox{\baselineskip12pt\hbox{UCSBTH-97-16}
\hbox{hep-th/9707102}
}}
{\vbox{\centerline {Comparing D-branes and Black Holes}
\vskip2pt\centerline{with 0- and 6-brane Charge}
}}

\centerline{{\ticp John M. Pierre}}
\vskip.1in
\centerline{\sl Department of Physics}
\centerline{\sl University of California}
\centerline{\sl Santa Barbara, CA 93106-9530}
\centerline{\sl jpierre@physics.ucsb.edu}
\bigskip
\bigskip
\centerline{\bf Abstract}
We consider configurations of D6-branes with D0-brane charge
given by recent work of Taylor
and compute interaction potentials with various D-brane probes
using a 1-loop open string calculation.
These results are compared to a supergravity calculation using
the solution
given by Sheinblatt of an extremal black hole carrying 0-brane and 6-brane
charge.  
\Date{}


\newsec{Introduction}

Among their many uses, D-branes \tasi\ can serve as probes 
of spacetime at large
distances as well as distances shorter than the string scale
\dkps \foot{For a review see \mrd}. 
Following the work of \SandV\ we have also learned that
configurations of D-branes can also serve as a microscopic 
description of black holes at weak 
coupling\foot{See \juanphd for a review and references}.
The fact that D-branes can be sensitive probes of
the structure of black holes at least at the one-loop level
has been shown in \refs{\dps,\juan}.  In order to more 
fully understand the correspondence between D-branes and
black holes, it is useful to study many examples, including
those with no supersymmetry.  Once in hand, a D-brane
description can be useful in understanding entropy, Hawking
radiation, and possibly the information problem.

In this short note we investigate the relationship between a
configuration of D6-branes that carry D0-brane charge
found in \wati\ and a black hole solution of IIA supergravity
that carries 0-brane and 6-brane charge studied in \hjs.
Since the static force between a 0-brane and 6-brane is 
repulsive at short and long distances, the conventional
wisdom has been that these objects cannot bind together.
However in \hjs\ it was argued that 0-6 
fermionic string modes could provide an attractive potential
and account for the Bekenstein-Hawking entropy of the
(metastable) black hole solution.  Further,
in \wati\ a configuration (not just a single 0-brane and 6-brane)
was given that carries only 0-brane and 6-brane charge and
is stable to quadratic order in fluctuations.
It is interesting to reconcile these two descriptions.
By examining their long distance interactions we find
evidence for their equivalence as was done in \lifsone\
for the case of black p-branes.
In section 2 we consider interactions of the 6-0-brane
``bound state'' with other D-brane probes using the 1-loop
annulus diagram in open string perturbation theory.  In
section 3 we compute the effective action of D-brane
probes in the background of the classical supergravity solution and we
find that the long distance interaction potentials agree in each case.

\newsec{Probing 6-branes in String Theory}
\seclab\strings
\subsec{The 6-0 Configuration}

We use the basic construction given in \refs{\wati} which can be
thought of as a configuration of four coincident D6-branes with
a world volume $U(4)$ gauge field strength,

\eqn\config{\eqalign{
F_{12} &= F\ {\rm diag}(1, 1, -1, -1) \cr
F_{34} &= F\ {\rm diag}(1, -1, -1, 1) \cr
F_{56} &= F\ {\rm diag}(1, -1, 1, -1) \cr}.}
This carries nonzero 0-brane charge 
($\int {\rm Tr}\ F\wedge F\wedge F \neq 0$),
with vanishing 2-brane 
($\int {\rm Tr}\ F\wedge F = 0$)
and 4-brane 
($\int {\rm Tr}\ F = 0$)
charge.  This gauge field configuration was shown to be stable classically
to quadratic order, but overall stability of this system is
an open question.
Since we take the world volume gauge field strength to be
diagonal, we only need consider the $U(1)^4$
subgroup of the gauge theory.  In order to compute the
one-loop scattering amplitude using open string perturbation
theory, we just take a composite of four D6-branes, each with
a constant background $U(1)$ field strength.

\subsec{Probing with 0-branes}

The one-loop
vacuum amplitude for open strings stretched
between a 0-brane in the presence of
6-brane with a constant background magnetic field strength
$F_{(2i-1)(2i)} = \tan\pi\epsilon_i$, 
relative velocity
$v = \tanh\pi\nu$,  and impact parameter $b$
is given by \refs{\lifsfour},

\eqn\ampzero{\eqalign{
{\cal A} &= {1\over 2\pi}\int {dt\over t}
e^{-b^2 t} Z_B \times Z_F \cr
Z_B &= {\Theta_1'(0|it) \over \Theta_1(\nu t|it)}
f_1^{-2}\thta{4}{1}^{-1}\thta{4}{2}^{-1}\thta{4}{3}^{-1} \cr
Z_F &= {1\over 2} \bigg\{ {\Theta_3(\nu t|it)\over \Theta_3(0|it)}
f_3^2 \thta{2}{1}\thta{2}{2}\thta{2}{3} \cr
&\quad-i{\Theta_4(\nu t|it)\over\Theta_4(0|it)}
f_4^2\thta{1}{1}\thta{1}{2}\thta{1}{3} \cr
&\quad-{\Theta_2(\nu t|it)\over\Theta_2(0|it)}
f_2^2\thta{3}{1}\thta{3}{2}\thta{3}{3}\bigg\} \cr}}
in units where $2\pi\alpha' = 1$.
To reproduce the 6-0 configuration we add four copies of
this amplitude each with 
$(\epsilon_1,\epsilon_2,\epsilon_3) = (\epsilon, \epsilon, \epsilon)$,
$(\epsilon, -\epsilon, -\epsilon)$, 
$(-\epsilon, -\epsilon, \epsilon)$, 
and $(-\epsilon, \epsilon, -\epsilon)$ respectively.
In this case however, each copy gives the same result.

In the $t\rightarrow \infty$ limit the amplitude becomes,
\eqn\shortzero{
{\cal A} \rightarrow -{1\over 2}\int {dt\over t} e^{-b^2 t} \cot vt.}
This is the same result as one gets from integrating out massive
modes of $0-6$ strings in the supersymmetric quantum mechanics of 
a D0-brane probe moving by a pure D6-brane which gives a determinant,

\eqn\SQM{\det \pmatrix{\partial_\tau & v t-ib \cr
vt+ib & \partial\tau \cr}.}
At short distances the 0-brane probe is not sensitive to a small
background magnetic field which represents lower
dimensional branes dissolved into the 6-brane.

By taking the $t\rightarrow 0$ limit we extract a long distance
potential due to exchange of massless closed strings which is
defined by 
${\cal A} = - \int^{+\infty}_{-\infty} d\tau V(r^2 = b^2 + \tau^2 v^2)$,
\eqn\VDzero{V = - 4 \times v
{\cosh 2\pi\nu - 3 \cos 2\pi\epsilon - 4 \cosh\pi\nu \sin^3\pi\epsilon
\over 8 \sinh\pi\nu \cos^3\pi\epsilon} r^{-1}}

\subsec{Probing with 2-branes}

Since 0-brane probes are not sensitive to differences between the
constituents which form the composite \config, it is useful to consider
other Dp-brane probes which are.  First consider a D2-brane probe
without any background gauge field turned on in its world volume.
Using techniques of \refs{\acny,\pair,\bachas,\lifsone,
\jabbari,\lifstwo}\ 
we find the one-loop amplitude is given by,

\eqn\amptwo{\eqalign{
{\cal A} &= {1\over 2\pi} V_2 \int {dt\over t}
e^{-b^2 t} Z_B \times Z_F \cr
Z_B &= {\Theta_1'(0|it) \over \Theta_1(\nu t|it)}
{i\tan\pi\epsilon_1\over 2\pi\thta{1}{1}}
f_1^{-2}\thta{4}{2}^{-1}\thta{4}{3}^{-1} \cr
Z_F &= {1\over 2} \bigg\{ {\Theta_3(\nu t|it)\over \Theta_3(0|it)}
f_3^2 \thta{3}{1}\thta{2}{2}\thta{2}{3} \cr
&\quad-{\Theta_4(\nu t|it)\over\Theta_4(0|it)}
f_4^2\thta{4}{1}\thta{1}{2}\thta{1}{3} \cr
&\quad-{\Theta_2(\nu t|it)\over\Theta_2(0|it)}
f_2^2\thta{2}{1}\thta{3}{2}\thta{3}{3}\bigg\} .\cr}}
In this case adding the four contributions to the amplitude
cancels the second term in $Z_F$ coming from the $(-1)^F$NS
sector.  In the $t\rightarrow\infty$ short distance limit the amplitude
becomes,

\eqn\shorttwo{
{\cal A} \rightarrow 
4\times{\tan\pi\epsilon \over 4\pi} \int {dt\over t} e^{-b^2 t} 
{\cosh\pi\epsilon t - \cos\pi\nu t \over \sin\pi\nu t \tanh \pi\epsilon t}.}
For small $v$ there is a tachyonic instability \antibrane\
at $b^2 < \pi \epsilon$.
The long distance potential in the $t\rightarrow 0$ limit is,
\eqn\VDtwo{V = - 4 \times v
{\cosh 2\pi\nu - \cos 2\pi\epsilon
\over 8 \sinh\pi\nu \cos^3\pi\epsilon} (2\pi)^{-1} r^{-1}.}

\subsec{Probing with 4-branes}

Next we consider a D4-brane probe.  The one-loop amplitude is,

\eqn\ampfour{\eqalign{
{\cal A} &= {1\over 2\pi} V_4 \int {dt\over t}
e^{-b^2 t} Z_B \times Z_F \cr
Z_B &= {\Theta_1'(0|it) \over \Theta_1(\nu t|it)}
{i\tan\pi\epsilon_1\over 2\pi\thta{1}{1}}
{i\tan\pi\epsilon_2\over 2\pi\thta{1}{2}}
f_1^{-2}\thta{4}{3}^{-1} \cr
Z_F &= {1\over 2} \bigg\{ {\Theta_3(\nu t|it)\over \Theta_3(0|it)}
f_3^2 \thta{3}{1}\thta{3}{2}\thta{2}{3} \cr
&\quad+i{\Theta_4(\nu t|it)\over\Theta_4(0|it)}
f_4^2\thta{4}{1}\thta{4}{2}\thta{1}{3} \cr
&\quad-{\Theta_2(\nu t|it)\over\Theta_2(0|it)}
f_2^2\thta{2}{1}\thta{2}{2}\thta{3}{3}\bigg\} .\cr}}
The calculation proceeds as in  the D2-brane case.
The $t\rightarrow\infty$ limit is,

\eqn\shortfour{{\cal A} \rightarrow 4 \times 
\bigg({\tan\pi\epsilon \over 4\pi}\bigg)^2 {1\over 2} 
\int {dt \over t} e^{-b^2 t} e^{\pi t \over 2}
{\cosh\pi\epsilon t \over \sin\pi\nu t \sinh^2\pi\epsilon t},}
and there is a tachyonic instability when
$b^2 < {\pi\over 2} - \pi\epsilon$.
The long distance ($t\rightarrow 0$) potential is,

\eqn\VDfour{V = - 4 \times v
{\cosh 2\pi\nu + \cos 2\pi\epsilon
\over 8 \sinh\pi\nu \cos^3\pi\epsilon} (2\pi)^{-2} r^{-1}.}

\newsec{Probing the Supergravity Black Hole Background}
\seclab\sugra
The supergravity solution for an extremal black hole with
0-brane and 6-brane charge given in \hjs\ is,

\eqn\sugraBH{\eqalign{
ds^2_{10} &= - H^2 dt^2 + H^{-2} dr^2
+ r^2 d\Omega^2 + dy_i dy^i \cr
H &= 1 - {q g \sqrt{\alpha'} \over r} \cr
C^{(1)} &= - \sqrt{2} g q 
\bigg[ {\sqrt{\alpha'}\over r}dt + (1-\cos\theta) d\phi\bigg] \cr
e^{2\phi} &= 1 \cr}.}
In this case with a constant dilaton there is a single parameter $q$
which is related to number of 6-branes by $q = {\sqrt{2}\over 4} Q_6$,
and further the number of 0-branes is fixed by 

\eqn\ratio{Q_0/Q_6 = V^6/(2\pi)^6 \alpha'^3.}

In this section we compute the effective action for various D-brane
probes \refs{\noforce,\cheptsey,\dps}
by expanding the bosonic action for a Dp-brane,

\eqn\Daction{S_p = - T_p \int d\tau d^p\sigma e^{-\phi}
\sqrt{-\det g_{\mu\nu} \partial X^\mu \partial X^\nu} 
+ T_p \int C^{(p+1)}}
around the background \sugraBH.  In each case we choose the 
static gauge,

\eqn\static{\eqalign{
X^0 &= \tau \cr
X^{1,..,p} &= \sigma^{1,..,p} \cr
X^{p+1,..,9} &= X^{p+1,..,9}(\tau) \cr}}
and consider only velocities transverse to the branes.  Expanding
the term with the square root to leading order in velocities gives,

\eqn\oneterm{- T_p V_p \int d\tau H +
{1\over 2}T_p V_p \int d\tau 
H^{-1} (H^{-2}{\dot r}^2 + r^2 {\dot \Omega}^2) }
As in \juan\ we must also make a change of variable,

\eqn\diffeq{{d\rho\over \rho} = {dr \over  r H}}
in order to bring the velocity into the standard form 
$v^2 = {\dot \rho}^2 + \rho^2 {\dot \Omega}^2$.  Only the case with
a 0-brane probe couples to the RR form $C^{(1)}$ through the second
term in \Daction\ which gives a contribution of
\foot{As in \lifsone\ we ignore the angular dependence 
to suppress the Lorentz force.}

\eqn\twoterm{ - T_0  V_p \int d\tau {g q \sqrt{2\alpha'} \over r}}
to the effective action in addition to \oneterm.  By expanding 
\oneterm\ and \twoterm\ to lowest order in ${1\over r}$ and using
$T_p = Q_p g^{-1} (2\pi)^{-p}(\alpha')^{-{p+1 \over 2}}$ 
we can read off the
leading term in the effective potential in each case and get the result,

\eqn\Vzero{V^{(0)}_{eff} \sim - Q_6 {\sqrt{2}\over 4}
[1 - \sqrt{2} + {3\over 2} v^2] \rho^{-1}}
for 0-brane probes and

\eqn\Vp{V^{(p)}_{eff} \sim - Q_6  {\sqrt{2}\over 4}
[1 + {3\over 2} v^2] (2\pi)^{-{p\over 2}}\rho^{-1}}
for 2-brane and 4-brane probes.

Comparing the supergravity results \Vzero, \Vp\ with the
string theory results \VDzero, \VDtwo, \VDfour\ we find precise
agreement for the static and $v^2$ terms if we set 
$\epsilon={1\over 4}$ which means $F=1$.  
Since $F$ is related to the relative D-brane charges through,

\eqn\Fterms{T_0 = T_p \int_{V^6} F^3}
using the D-brane tension formula, we get the same relation 
on the charges as \ratio.

\newsec{Discussion}

In this paper we have considered the interactions of D-brane probes
in two very different contexts.  In one case we scattered probes off a 
D-brane configuration carrying only 0- and 6-brane charge using
open string perturbation theory, and in the
other case we considered the motion of these probes in the background
of a classical black hole solution.  As was speculated in \wati\
we have found that at large distances these two objects look alike.
In \hjs\ is was argued that the supergravity solution is metastable
only for large values of $Q_6$ and $Q_0$, however the probe 
calculation considered here seems to hold for smaller values of
the charges as well.

The 6-0 D-brane configuration also has a description within
the context of M(atrix) theory and it was recently considered
in \VandK.  There it was found that M(atrix) theory gave the
correct result compared with a 6-brane probe carrying a very large
amount of 0-brane charge.  Here, since the ratio of
0-brane and 6-brane charge is fixed by the supergravity solution
with a constant dilaton, we cannot arbitrarily increase the amount
of relative 0-brane charge and go to a regime where light open
string modes (and hence M(atrix)
theory) agree with the classical result.
It would be interesting to consider the generalization of \sugraBH\
to the case with a non-constant dilaton where 
this ratio is a free parameter and to
study the regime where a short distance open string description
is valid at long distances.



\bigskip\bigskip\centerline{{\bf Acknowledgements}}\nobreak

I am grateful to S. Giddings,
J. Polchinski, and H. Sheinblatt for helpful discussions.  
I would also like to thank G. Lifschytz,
J. Maldacena, and W. Taylor for
correspondences.
This work is supported in part by DOE grant
DOE-91ER40618 and NSF PYI grant PHY-9157463.

\listrefs

\end